# Audio-Visual Biometric Recognition and Presentation Attack Detection: A Comprehensive Survey

HAREESH MANDALAPU[1], ARAVINDA REDDY P N[2],
RAGHAVENDRA RAMACHANDRA[1], (Senior Member, IEEE),
KROTHAPALLI SREENIVASA RAO[3], (Member, IEEE),
PABITRA MITRA[3], (Member, IEEE),
S. R. MAHADEVA PRASANNA[4], (Member, IEEE),
AND CHRISTOPH BUSCH[1], (Senior Member, IEEE)

[1]Department of Information Security and Communication Technology, Norwegian University of Science and Technology (NTNU), 2815 Gjøvik, Norway
[2]Advanced Technology Development Centre, Indian Institute of Technology Kharagpur, Kharagpur 721302, India
[3]Department of Computer Science and Engineering, Indian Institute of Technology Kharagpur, Kharagpur 721302, India
[4]Department of Electrical Engineering, Indian Institute of Technology Dharwad, Dharwad 580011, India

Corresponding author: Hareesh Mandalapu (hareesh.mandalapu@ntnu.no)

This work was supported by Department of Information Security and Communication Technology, NTNU, Gjøvik and Advanced Technology Development Centre, Indian Institute of Technology, Kharagpur, India.

**ABSTRACT** Biometric recognition is a trending technology that uses unique characteristics data to identify or verify/authenticate security applications. Amidst the classically used biometrics, voice and face attributes are the most propitious for prevalent applications in day-to-day life because they are easy to obtain through restrained and user-friendly procedures. The pervasiveness of low-cost audio and face capture sensors in smartphones, laptops, and tablets has made the advantage of voice and face biometrics more exceptional when compared to other biometrics. For many years, acoustic information alone has been a great success in automatic speaker verification applications. Meantime, the last decade or two has also witnessed a remarkable ascent in face recognition technologies. Nonetheless, in adverse unconstrained environments, neither of these techniques achieves optimal performance. Since audio-visual information carries correlated and complementary information, integrating them into one recognition system can increase the system's performance. The vulnerability of biometrics towards presentation attacks and audio-visual data usage for the detection of such attacks is also a hot topic of research. This paper made a comprehensive survey on existing state-of-the-art audio-visual recognition techniques, publicly available databases for benchmarking, and Presentation Attack Detection (PAD) algorithms. Further, a detailed discussion on challenges and open problems is presented in this field of biometrics.

**INDEX TERMS** Biometrics, audio-visual person recognition, presentation attack detection.

## I. INTRODUCTION

Biometric technology is swiftly gaining popularity and has become a crucial part of day-to-day life. A biometric system aims to recognize a data subject based on their physiological or behavioral characteristics [8]. Recognition systems are based on biometric characteristics, such as DNA, face, iris, finger vein, fingerprint, keystroke, voice, and gait. Several factors are considered while designing and applying biometrics: accuracy to authentication, robustness to spoof or impostor attacks, user acceptance, and cost of capture sensors. Amidst these factors, user acceptance and sensor cost are the primary hindrances that thwart highly accurate and robust biometrics.

The authentication system that uses a single biometric cue such as speech or face is called a unimodal system. The biometric cue can use more than one classifier and employ a fusion approach to perform recognition. Nonetheless, the captured biometric cue may be of low quality due to variations in pose, illuminations, background noise, and low spatial and temporal resolution of the video. This problem is addressed by using multiple biometric modalities for

The associate editor coordinating the review of this manuscript and approving it for publication was Sedat Akleylek[ID].

authentication [113]. Deploying multimodal data introduces other problems like multiple captures, processing time, and design overhead. The vulnerabilities present by unimodal biometrics may also exist in a multimodal system. The audio-visual biometrics took multimodal biometrics to another better level by taking advantage of complimentary biometric information present between voice and face cues. In analogy, voice and face biometrics are most user-friendly and cost-effective as they allow capturing the multi-biometrics in a single capture using low-cost sensors (e. g., smartphone camera). These points made audio-visual biometrics an exciting topic of research in field of multimodal biometrics.

Audio-visual biometrics has gained interest among biometric researchers both in academics and in industry. As a result there are an ample amount of literature available [8], [72], [78], [126], publicly available databases [12], [24], [25], [85], [102], [118], [137], devoted books [21], open-source software [10], [104], mobile applications [54], [127], speaker and recognition competitions [74], [116]. The National Institute of Standards and Technology (NIST) conducted a challenge of Audio-Visual speaker recognition in 2019 (Audio-visual SRE19) [116]. The challenge provided baseline face recognition and speaker recognition and accepted two evaluation tracks, audio-only and audio-visual, along with visual-only as an optional track. This competition submissions have indicated interesting results and started a new direction in audio-visual biometrics in ongoing NIST SRE challenges. Further, there is an ongoing multimodal biometric project called RESPECT [110] which is on the verge of producing a robust audio-visual biometric recognition system. As an application, there are smartphones for performing financial transactions (e. g. banking transactions, Google pay, e-government, e-commerce), border control [42] where AV biometrics can be deployed because they provide an ideal choice for subdued and low-cost automatic recognition. Although there are no commercial biometric systems that use only audio-visual person authentication, there are domains where multimodal biometrics are used. The dependency of the constrained environment for audio-visual data capture limits the commercial use of AV biometrics. However, looking at smartphones usage growth, which is equipped with high-quality cameras and microphones, there is a scope to use audio-visual biometrics in real-world applications.

In this survey paper, we discuss audio-visual (AV) biometrics, where speech is used along with stagnant video frames of the face or certain parts of the face [15], [23], [33], [59], [120] or video frames of the face or mouth region (visual speech) [27], [40], [70], [139], [140] in order to improve the performance. The face and speech traits are fused either at the feature level (i,e., features are fused and fed to the classifier) or at the score level (i,e., an individual recognition system is built for each trait, and scores from the system are fused). We have discussed different types of fusion schemes used in AV biometrics. The main goal of audio-visual biometrics is to improve the robustness of recognition towards unconstrained conditions and vulnerabilities. Biometric attributes (face and speech) are prone to presentation attacks where a unimodal system produces dubious recognition results. This paper also presents several presentation attack detection (PAD) algorithms that used complimentary audio-visual information (over a single cue) to obtain robust biometric systems [20], [29], [30], [71], [114].

Few survey papers are available in the literature to provide a concise review of audio-visual biometrics, including feature extraction, speaker recognition process, fusion methods, and AV databases. Deravi [39] has reviewed the audio-visual biometric systems in application to access control. Aleksic *et al.* [8] presented a survey of audio-visual biometric methods, fusion approaches, and databases until 2006. Li has performed a survey on authentication methods based on audio-visual biometrics [78] with brief reviews and presented a comparison of audio-visual biometrics until 2012. The existing papers have also discussed some of the audio-visual biometric systems [7], [15], [48], [89], [120], [144] that are vulnerable to replay attacks. There are survey papers only on the fusion approaches used in AV biometric data fusion [31], [126]. This survey paper presents a thorough review of all spearhead efforts in AV biometrics and presentation attack detection (PAD) algorithms.

By considering the above survey papers and emerged technologies in AV biometrics, this work contributes to the following:

1) A complete up to date review of existing AV biometric systems and detailed discussion on audio-visual databases.
2) A detailed description of different audio and visual features, fusion approaches, and achieved performances are presented.
3) A thorough review of existing presentation attack detection (PAD) algorithms for audio-visual biometrics is performed.
4) Challenges and drawbacks, emerging problems, and privacy-preserving techniques in audio-visual biometrics are presented.

The rest of the paper is organized as follows: Section II presents the general concepts of AV biometric recognition system, and section III presents the features in AV biometrics. In section IV, we present different approaches used in audio-visual fusion and classification. Section V discusses the existing audio-visual databases and the comparison of benchmark AV algorithms on each database. Further, section VI describes PAD algorithms on AV based biometrics. Section VII presents challenges and open questions in this research domain and we conclude the report in section VIII along with discussion of future works in this direction.

## II. GENERAL CONCEPTS OF AV BIOMETRIC VERIFICATION SYSTEM

This section discusses different types of audio-visual biometric systems and ISO Standard (ISO/IEC JTC1 SC37

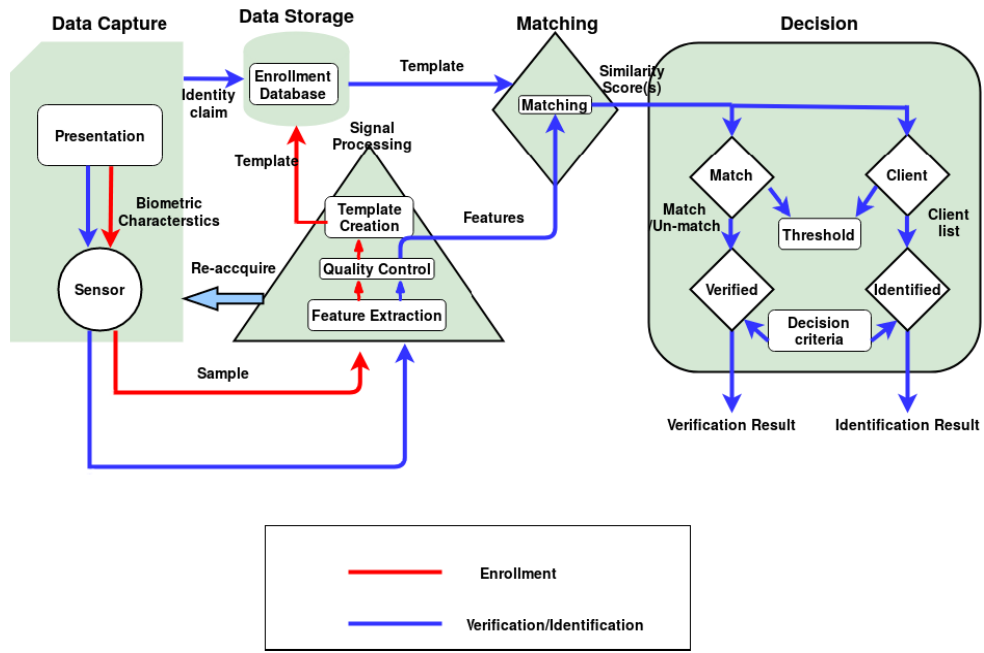

**FIGURE 1.** Conceptual biometric model inspired from ISO/IEC JTC1 SC37.

Biometrics 2016) [66] biometric components. An AV biometric recognition can be classified into two types: identification and verification. Identification is a process of finding out an individual's identity by comparing the biometric sample collected from the subject with all the individuals from the database. Verification is a process where the claimed identity is checked against a single model where the biometric sample collected from an individual is compared with the same individual's sample from the database. The AV biometric system can also be divided into two types based on audio and visual data captured. Depending on the text uttered by the speaker, the AV biometric system can be called a Text-dependent or a Text-independent system. If the AV biometric system uses static visual information (e.g., an image of a face or static faces from video frames) is called Audio-Visual-Static biometric systems. In contrast, AV systems using visual features containing temporal information from video frames are called Audio-Visual-Dynamic biometric systems.

### A. BIOMETRIC SYSTEM COMPONENTS

Figure 1 shows a block diagram of ISO/IEC JTC1 SC37 biometrics recognition diagram [66] describing two main phases of a biometrics system namely, enrollment phase (red-colored lines) and verification or identification phase (blue colored line). There are five major stages in this system: data capture, Signal processing, Data storage, Matching, and Decision making, as indicated in [66]. Data capture, signal processing, and Data Storage are used only in enrollment, and the rest of the blocks are used in both enrollment and recognition phases. The first stage is the data capture, where audio-visual biometric is captured using a sensor and the second stage is the signal processing block, which includes multiple steps. For example, segmentation and feature extraction are carried out in this step by cropping out the biometric region and extracting optimal features.

Pre-processing is a part of the signal processing block where the biometric sample is prepared for feature extraction. Pre-processing of an audio signal includes signal denoising [89], channel noise removal, smoothing [107], signal enhancement, silence detection and removal. Pre-processing a video signal consists of steps like detecting and tracking the face or any other important face regions. After feature extraction, the next sub-block in the signal processing stage is the biometric sample's quality control. A biometric sample is of acceptable quality if it is suitable for person recognition. According to ISO/IEC 29794-1 standardization [67], we have established three components of a biometric sample, namely *Character:* implies the source's built-in discriminative capability, *Fidelity:* the degree of resemblance between a sample and the source, and *Utility:* the samples' impact on biometric systems' all-around performance.

The next sub-block is the Data Storage, where a biometric template is created. A biometric template is a digital footnote of the peculiar characteristics of a biometric sample. Created templates are stored in the database and are used at the time of authentication. Once the sample biometric digital reference is stored in the database in the enrollment phase, the digital footnote is matched with the person seeking authentication or identification, and a binary decision, accept or reject, is made based upon a threshold both in the identification and in verification.

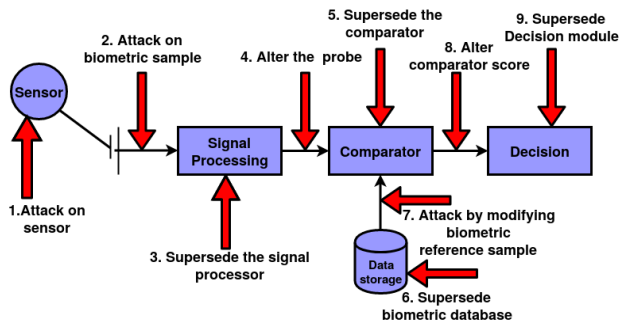

**FIGURE 2.** Vulnerability of AV biometric system (motivated by figure ISO/IEC 30107-1).

### B. PRESENTATION ATTACK DETECTION (PAD)

A biometric recognition system is prone to multiple types of threats. Among these, presentation attacks are considered to be one of the significant vulnerabilities. Figure 2 shows the generic block diagram of the biometric recognition system (in our case, audio-visual) with nine contrasting vulnerabilities, as illustrated in ISO/IEC 30107-1 [68]. The first vulnerability is at the sensor, where a pre-recorded audio or face image artifact of a lawful client is presented as an input to the sensor. An artifact as defined in ISO/IEC JTC1 SC37 Biometrics 2016 [68] is a morphed object or depiction presenting a copy of biometric characteristics or fabricated biometric patterns. This kind of attack is also known as a presentation attack.

Presentation attacks are defined as the presentation to a biometric capture subsystem with the goal of interfering with the operation of the biometric system [68]. Presentation Attack Instrument (PAI) is the biometric characteristic or the object used in a presentation attack. Presentation attacks can be divided into two types: an active impostor presentation attack and a concealer presentation attack. The active impostor attacks are a type of attack in which the attacker tends to be recognized as a different subject. This type is again divided into two types. The first type is that the intention is to get recognized as a subject known to the AV biometric system. The second type is to get recognized as the unknown person to the AV biometric system. A concealer presentation attack is the type of attack where the subject tries to avoid getting recognized as a subject in the system.

The popular presentation attacks in audio-visual biometrics are replay attacks and forgery attacks. A replay attack is performed by replaying the audio-visual recording sample in front of a biometric sensor. This can be performed either on individual modality (face video replay or audio recording replay) or both modalities at once (audio-visual replay). Forgery attack is carried out by altering the audio-visual sample to make it look like a bona fide sample of the target speaker. Audio-visual forgery includes two modal transformations. Speaker transformation, also known as voice transformation, voice conversion, or speaker forgery, is a technique for altering an impostor's utterance to make it sound like the target speaker (client). In the visual domain, face transformation aims at creating an animated face synthetically from a still image of the target client.

Presentation Attack Detection (PAD) is a framework by which presentation attacks can be identified to be classified, particularised, and communicated for decision-making and performance analysis. In literature, PAD is also termed as anti-spoofing techniques in the development of countermeasures to the biometric spoofs. In most of the existing AV biometrics literature, PAD is referred to as liveness detection; however, liveness detection is defined as the measurement and analysis of involuntary or voluntary reactions in order to detect and verify whether or not a biometric modality presented is alive from a subject at the time of capture [68]. So, from the standardization, we can infer that liveness detection is considered a subset of PAD but not as a synonym for itself.

### C. PERFORMANCE METRICS

In this section, we discuss the performance metrics used in the field of audio-visual biometric methods.

False Match Rate (FMR) is the percentage of impostors samples accepted by the biometric algorithm, and False Non-Match Rate (FNMR) is the percentage of bona fide samples rejected by the algorithm [66]. At a biometric system-level performance, Fale Acceptance Rate (FAR) and False Rejection Rate (FRR) are reported in the place of FMR and FNMR, respectively. Many research works used an equal error rate (EER) to represent FMR and FNMR metrics in a single value. EER is a single value at which FMR and FNMR are equal. Similarly, the Total Error Rate (TER) is the sum of FAR and FRR, and half TER (HTER) is the average of the FAR and FRR. Some algorithms mentioned the accuracy rate or error rate, which is the percentage of samples being correctly classified or incorrectly classified, respectively.

## III. AV BASED FEATURE EXTRACTION

This section presents a brief overview of the AV features widely employed in designing the multimodal biometric system based on face and voice. Features are the distinct properties of the input signal that helps in making a distinction between biometric samples. Feature extraction can be defined as transforming the input signal into a limited set of values. Further, feature extraction is useful to discard extraneous information without losing relevant information. The majority of the literature has treated AV biometrics as two unimodal biometrics based on visual (or face) and audio (or voice) biometric characteristics. Thus, the feature extraction techniques are carried out independently on audio and visual biometrics that are briefly discussed below.

### A. AUDIO FEATURES

The Audio features used in audio-visual biometric methods are classified into four categories, as depicted in Figure 3. The details of various types of audio features are briefly discussed in the following subsections.

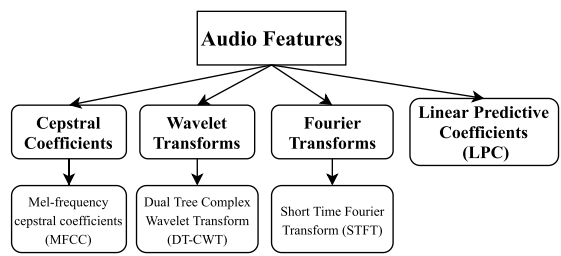

**FIGURE 3.** Different types of audio features used for audio-visual biometric recognition.

### 1) CEPSTRAL COEFFICIENTS

The cepstrum of a signal is obtained by applying an inverse Fourier transform of the logarithm. The logarithm is calculated from the magnitude of the Fourier transform. The advantages of cepstrum include its robustness and separation of excitation source and vocal tract system features. Robert *et al.* [50] from dialog communication systems developed a multimodal identification system where speech utterance is divided into several overlapping frames, and cepstral coefficients are extracted from these frames and used as features for AV biometric recognition.

Among cepstral coefficients, Mel-frequency cepstral coefficients (MFCC) representation is an efficient speech feature based on human auditory perceptions. MFCCs include series of operations, namely pre-emphasis (increasing magnitude of higher frequencies), framing (speech signal is divided into chunks by a window), applying Fast Fourier Transform (FFT), Mel-filtering, followed by applying DCT on log filter banks (where lower-order coefficients represent vocal tract information) to obtain the MFCCs. Mel-frequency banks approximate the human ear response more accurately than any other system, and MFCCs suppress the minor spectral variations in higher frequency bands.

MFCCs have been widely used AV for person recognition [6], [7], [16], [23], [35], [46], [69], [73], [89], [94], [123], [125], [144], [145]. Classification methods based on Gaussian Mixture Models (GMMs), Vector Quantization (VQ) have displayed a consistent speaker recognition performance using MFCCs. Experiments conducted on XM2VTS database [88], AMP/CMU database [80], VidTIMIT [117], [118] have displayed robustness of MFCCs in accurate person identification. Mobile applications [90], [132] have also used MFCCs as feature vectors. Neural network based methods [65] have examined cepstral coefficients namely i) Real Cepstral Coefficients (RCCs), ii) Linear Prediction Cepstral Coefficients (LPCC), iii) MFCCs, iv) $\Delta$MFCCs, and v) $\Delta\Delta$MFCCs. It is observed that $\Delta$MFCCs have performed better than others. Alam *et al.* [4], [5] have explored the usage of MFCCs in deep neural network based methods. Further, MFCCs are also used in creating i-vectors, which performed better with Linear Discriminant Analysis (LDA) and Within Class Covariance Normalisation (WCCN) [105]. In the recent works, MFCCs are used as a potential complementing feature in multimodal biometrics [53], [87].

### 2) WAVELET TRANSFORMS

The popular wavelet transform approach used in speaker recognition methods is the Dual tree complex wavelet transform (DTCWT). DTCWT uses two discrete wavelet transforms (DWT) in parallel [121], one DTCWT generates a real part of the signal other DTCWT generates the imaginary part. DTCWT is highly directional, shift-invariant, offers perfect reconstruction, and computationally efficient. Another variant of the wavelet transform is the Dual-Tree Complex Wavelet Packet Transform (DT-CWPT) [143]. It is observed that using DT-CWPT has increased the speaker identification rates in both unimodal and multimodal systems when compared to MFCCs based methods.

### 3) FOURIER TRANSFORMS

The Short-time Fourier transform (STFT) is a popular Fourier transform approach used in the processing of voice as biometric data. The voice signal is a quasi-stationary signal; therefore, STFT yields better representation over a Fourier transform. In STFT, the speech utterance is segmented into frames of a smaller duration, approximately 20-30ms, and Hamming or Hanning window is superimposed on these frames before computing the Fourier transforms. The window slides throughout the signal with an overlap between the frames. Dieckmann *et al.* [40] presented a Synergetic Computer-based biometric identification system where a Hanning window covers the input signal, and STFT is applied. A power function is applied to emphasize the lower frequencies and to compress the higher frequencies.

### 4) LINEAR PREDICTION COEFFICIENTS(LPC)

The continuous-time speech signal is highly correlated. If we know the previous sample, it is possible to predict the next sample. The linear predictor predicts the next point as a linear combination of previous values. The transfer function of a linear prediction filter is an all-pole model. Linear Prediction Coefficients (LPC) model the human vocal tract as a source-filter model. Here the source is the train of impulses generated by the vibration of vocal folds, which acts as an excitation source. The filter represents the oral cavity, which models the vocal tract system, and the resulting speech signal is the convolution of a train of impulses and responses of the vocal tract system. LPCs are a compact representation of the vocal tract system and can be used for synthesizing the speech. LPCs are used for deriving the LP residual (equivalent to excitation source) with inverse filter (all-zero filter) formulation. LPCs are used for speaker recognition in AV biometric methods [1], [14], [15], [41] using Hidden Markov Models (HMMs) and Gaussian Mixture Models (GMMs) for classification.

### B. VISUAL FEATURES

This section presents a brief overview of the visual (or facial) features that are classified into four major types, as shown in figure 4.

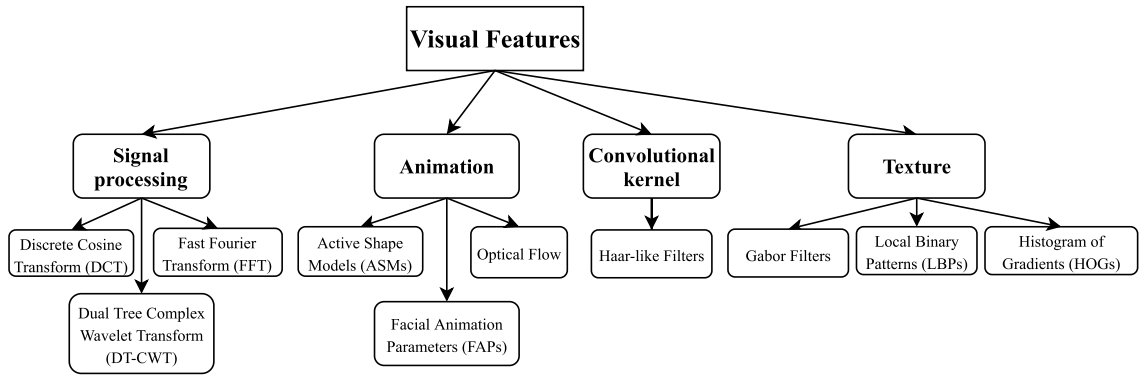

**FIGURE 4.** Different visual features used in audio-visual biometric recognition.

### 1) SIGNAL PROCESSING BASED FEATURE EXTRACTION

In signal processing based feature extraction, there are three different methods used in AV speaker recognition, namely Discrete cosine transform (DCT), Discrete-Time Complex Wavelet Transform (DT-CWT) iii) Fast Fourier Transform (FFT).

The discrete cosine transform (DCT) of an image represents the sum of sinusoids of varying frequencies and magnitudes. DCT has an inherent property that contains information about the image in the first few coefficients, and the rest can be discarded. DCT contains AC and DC coefficients where DC coefficients are prone to illumination changes, and hence they are discarded. However, first, few AC coefficients act as a good representation of an image. Therefore, DCTs are widely used in feature extraction and compression techniques. In the early works on audio-visual fusion for biometrics, DCTs are computed on small blocks of the image [123] and appended with the mean and variance of overlapping blocks [90]. Further, four variants of DCT methods namely DCT-delta, DCT-mod, DCT-mod-delta and DCT-mod2 [119] are examined. DCT-mod2 is formed by replacing the first three coefficients of 2D-DCT with their delta coefficients and used as feature vectors [73].

Dual tree complex wavelet transforms (DTCWT) is another feature extraction approach used for face images similar to audio features described in Section III-A2. DTCWT features are extracted at different depths and convolved to form a feature vector by concatenating all the rows and columns [143]. To reduce these feature vectors' dimensionality, PCA is applied, and only 24 vectors are chosen from 6 directions. Using Fast Fourier Transform (FFT), an image can be transformed into the frequency domain as a sum of complex sinusoids with varying magnitudes, frequencies, and phases. The advantage of using FFT is that the $N$ transformed points can be expressed as a sum of $N/2$ points (divide and conquer), and thus, computations can be reused. Therefore, FFTs can be used for efficient feature extraction methods for texture analysis. Robert W *et al.* [50] developed a novel multimodal identification system for face recognition from videos where 3D FFTs of 16 vector fields are computed with unique identifiable points from lips and faces.

### 2) ANIMATION BASED FEATURES

The animation based visual features used in AV biometrics are active shape models (ASMs), facial animation parameters (FAPs), and optical flow features.

Active Shape Models (ASMs) are the statistical models of shape and appearance used to represent the face region in an image. Human experts annotate face images, and then a model is trained using a set of images. The ASM algorithm makes few postulates about the objects being modeled other than what it learns from the training set. ASMs not only give compact delineation of allowable variation but also avoid the unacceptable shapes being generated. ASMs are used to detect faces in images, and a neural network based AV speaker identification is employed in [65]. After successful face detection, region of interest is segmented using robust real-time skin color blob detection and radial scanline detection methods. Further, the background noise is eliminated, and finally, appearance-based face features are obtained [61]. Similarly, Bengio *et al.* [16] used point distribution models to track and extract the visual information from each image. For each image, 12 lip contour features and 12 intensity features, including their first-order derivatives, are excerpted, making a total of 48 features. Brunelli *et al.* [23] used pixel-level information from eyes, nose, and mouth regions to extract the features.

Facial animation parameters (FAPs) are a type of high-level features extracted from the lip-contour region. These high-level features have several advantages over low-level features like sensitivity to light and rotation. A 10-dimensional FAPs describing lip contours are extracted in [7], projected onto eigenspace to use in audio-visual person identification.

Optical flow is a probable motion of individual pixels on an image plane. The optical flow of the pixels can be computed by assuming Spatio-temporal variations in the image. Using a Charge Coupled Display (CCD) camera and infrared camera [40], horizontal and vertical projections of an image are computed and concatenated to a resulting gray level image and optical flow of mouth region [62]. A real-time face tracking and depth information is used to detect and recognize the face under varying pose in [35]. A dense optical flow algorithm is used to calculate the velocity of moving pixels and edges for AV person authentication in [46].

### 3) CONVOLUTION KERNEL BASED FEATURES

Well-known object convolution kernel methods are Haar-like filters that can detect edges and lines in an image effectively. Voila-Jones face detection algorithm [134] used Haar wavelets to detect the most relevant features from a face such as eyes, nose, lips, and forehead. Therefore, Haar-like features are extended for the application of face recognition [94]. Visual speech features are derived from the mouth region by cascaded algorithm portrayed in [79]. Similarly, the Viola-Jones algorithm is also used for successful face recognition [69]. Asymboost is an another efficient face detection algorithm that uses a multi-layer cascade classifier to detect the face in multiple poses [135]. Under different illuminations and non-cooperative situations like pose and occlusions, face recognition is a challenging task. Therefore, histogram equalization is performed to normalize the images after the image acquisition [89]. When the face is more occluded, Haar cascade classifiers are used for detecting the eye portion of the image. An integral image representation that reduces time complexity and uses Haar-based features to perform AV person identification in [6]. Further, K-SVD (Single Value Decomposition) algorithm is used to create a dictionary for every video sample [105] by taking advantage of high redundancy between the video frames. K-SVD is an efficient algorithm for adapting dictionaries to achieve sparse signal representations of faces detected in each frame [2].

### 4) TEXTURE BASED FEATURES

There are three types of texture-based features used in AV biometric methods, namely, Gabor filters, Local Binary Patterns (LBP), and Histogram of Gradients (HOG).

A Gabor filter is a sinusoidal signal with a given frequency and orientations modulated by Gaussians [146]. Since Gabor filters have orientation characteristics, they are extensively used in texture analysis and feature extraction of face images. Initial works in AV biometrics spotted face image by using best fitting the ellipse followed by identifying eyes and mouth position by topographic grey relief [1]. After successful face recognition, Gabor filters are applied to extract the features, and complex Gabor responses from filters with six orientation and three resolutions are used as feature vectors [41]. Machine learning algorithms like Support Vector Machines (SVM) with Elastic Graph Matching (EGM) have displayed noticeable results [14], [15]. Further, the Pyramidal Gabor-Eigenface algorithm (PGE) is used to extract the Gabor features [64], [144].

Local Binary Pattern (LBP) is a textual operator that labels the pixels in an image by considering the neighboring pixels' values and assigns a binary number. LBP for a center pixel is calculated first using the window and is binarised according to whether pixels have high value than the center pixel. LBP histogram is computed over the LBP output array. For a block, one of the $2^8 = 256$ possible patterns is possible. LBP's advantages include high discriminate power, computational simplicity, and invariant to gray-scale changes. The use of LBPs has shown a prominent advantage in face recognition approaches. LBPs features are used for face recognition using a semi-supervised discriminant analysis as an extension to linear discriminant analysis (LDA) [145]. Face regions in an image are detected by localizing lip and eye regions using Hough transforms [51], [134]. LBP features are extracted on the detected faces for multimodal authentication in [124], [125]. Deep neural network based AV recognition systems [4] employed LBPs as visual features from face images that are photometrically normalized using the Tan-Triggs algorithm [128]. In further research, a joint deep Boltzmann machine (jDBM) model that uses LBPs is introduced with an improved performance [5]. The histograms of the face and non-face region using LBP features are extracted, and a biometric classifier is implemented using pattern recognition in [101], [132].

Histogram of Gradients (HOG) is another popular texture feature descriptor used to extract robust features from images [36]. HOG features are chosen over Local Binary Patterns (LBP), Gabor filters, Scale Invariant Transform (SIFT) because of the properties like robustness to scale and rotation variance, and global features. The multimodal biometrics method used HOG via Discriminant Correlation Analysis (DCA) on mobile devices [53].

Table 1 shows how different audio-visual features are discussed in this survey.

## IV. AV BASED FUSION AND CLASSIFICATION

Information fusion is used to assimilate two complementary modalities with an eventual objective of attaining the best classification results. The audio-visual biometric methods have utilized many fusion approaches to complement audio and video characteristics to one another. The Figure 5 shows classification of audio-visual fusion methods. Fusion methods are divided mainly into three types: Pre-mapping (early fusion), Midst-mapping (intermediate fusion), and Post-mapping (late fusion). In this section, different audio-visual biometric methods are described with their corresponding performances. The methods described here include the performance of recognition without presentation attacks, i.e., the impostors are zero-effort impostors. The presentation attack detection algorithms used in audio-visual biometrics are discussed in Section VI.

### A. PRE-MAPPING OR EARLY FUSION

In the pre-mapping or early fusion approach, individual features from voice and face are fused to make a single set of features.

TABLE 1. Different audio and visual features used in AV biometric methods.

| Types of visual features | Types of audio features | | | |
|---|---|---|---|---|
| | Cepstral Coefficients (MFCC) | Wavelet Transforms (DTCWT) | Fourier Transforms (STFT) | Linear Prediction Coefficients (LPC) |
| Signal Processing | [123], [90], [73] | [143] | - | [50] |
| Animation | [65], [35], [46], [23] | - | [40], [16] | [7] |
| Convolutional Kernel | [94], [69], [89], [124], [6], [105] | - | - | - |
| Texture | [145], [125], [4], [5], [132], [53] | - | - | [1], [41], [14], [15] |

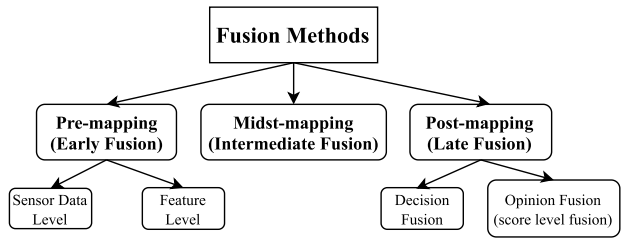

FIGURE 5. Audio-Visual fusion methods inspired from [120].

The earliest methods to use the pre-mapping of AV biometrics used fusion of static and dynamic features and used classifiers like a synergistic computer with MELT[1] algorithm [40]. The concept of synergistic computer SESAM[2] utilizes the combination of static and dynamic biometric characteristics, thus making the recognition system robust to imposters and criminal attacks. Further works explored Hidden Markov Models (HMM), which are trained using fused audio and visual features [7]. Gaussian mixture models (GMM) are also used as classifiers because of their low memory and well suitability for text-dependent and text-independent applications. GMM based classifications on concatenated features of audio and visual domains have displayed better performance than the score-level fusion [46], [123]. Some early fusion methods used clustering algorithms and PCA to reduce the dimensionality of features for efficient fusion [143]. As the cluster size is increased from 32 to 64, a higher identification rate is observed.

Laplacian projection matrix is another effective way of representing audio, and video features used in early fusion technique [69]. Laplacian Eigenmap [60] is an efficient nonlinear approach that can preserve inherent geometric data and local structure. The Laplacian matrices from both traits are fused linearly to form a single vector for audio-visual person recognition. Experiments conducted with pose estimation show an error rate of 35%. Without pose estimation, the error rate was 50%. The Laplacian Eigenmap fusion method outperforms the low-level fusion latent semantic analysis.

Multi-view Semi-Supervised Discriminant Analysis (MSDA) is an extension to Semi-supervised Discriminant Analysis (SDA) for feature level fusion [145]. The MSDA is inspired by a multi-view semi-supervised learning method called co-training [19]. A GMM mean adapted super vector and an LBP super vector is fused and fed into MSDA, PCA, Locality preserving projection (LPP), Linear discrimination analysis (LDA), and SDA individually. However, MSDA outperforms all other techniques because of local adjacency constraints, which can be effectively learned in different views using the same data. The synchronous measurement between audio and visual domains is examined in other works [144]. Synchronized feature vectors of size 21 are concatenated and fed to Probabilistic Neural Network (PNN). Experiments are performed at different resolutions of the face and different audio lengths (25s, 20s for training and 12.5s and 10s for testing). It is observed that the PNN method overcomes the difficulty of different frame rates for audio and visual signals and also the curse of dimensionality.

The time series vectors of face and speech provide unique characteristics of a person. The distance between data with different vector lengths is obtained using Dynamic Time Warping (DTW) [125] using the time series information from a video. The similarities between voice and face features are calculated using DTW, and multiple classifiers are fed with the similarity measures. Experiments show an authentication error of 0% for different kinds of clients. Feature-level fusion methods employed Quadratic Discriminant Analysis (QDA) for minimizing the misclassification rate, and an EER of 0.5% is obtained with the least memory and time consumption [132]. A Joint Deep Boltzmann Machine (jDBM) with a pre-training strategy and a joint restricted Boltzmann machine (jRBM) are used to model speech and face separately [5]. Then fused features were evaluated with JPEG compression and babble noise to degrade the face and speech files, respectively. The jDBM method outperforms bimodal DBM in significantly degrading conditions.

Decision voting is used for 39-dimensional audio, and video features [105]. During fusion, the standard sparsity concentration index was modified because face and speech cues are two complementary modalities and a new classification rule was derived called a joint sparse classifier. The proposed classifier outperforms the sparse representation classifier, which was used for a single modality. Discriminant Co-relation Analysis (DCA) is used to perform an early fusion of MFCCs and HOG features [53].

---

[1]MELT: the prototypes of one class are *melted* into one prototype
[2]Synergetische Erkennungmittels Standbild, Akustik und Motorik

The DCA fused feature set is given to five different classifiers, namely Support Vector Machine, Linear Discriminant Analysis, Quadratic Discriminant Analysis, Random Forests, and K-Nearest Neighbours. SVM achieves the lowest EER of 20.59% among all classifiers mentioned, and SVM requires 50.9818*s* for training and 0.6038*s* for testing, which is significantly less when compared to other classifiers.

### B. MIDST-MAPPING OR INTERMEDIATE FUSION

The midst-mapping or intermediate fusion is a relatively complicated technique compared to the early fusion technique. In this approach, several information streams are processed while mapping from the feature space to the decision space. The intermediate fusion technique exploits the temporal synchrony between the video streams (e.g., speech signal and lip movements of videos) by which the curse of dimensionality problems with feature level fusion technique can be avoided. Examples of this type of fusion are HMMs, which can handle multiple streams of data. Asynchronous HMMs are used for text-dependent multimodal authentication in [16]. Training of AHMMs was performed using the Expectation Maximisation (EM) algorithm with clean data. Experiments were conducted on AV samples with various noise levels (0dB, 5dB, 10dB), and results display promising Half Total Error Rate (HTER) compared to audio-only and face-only modalities. In the next works, Coupled Hidden Markov Models (CHMM) are used for audio recognition and Embedded Hidden Markov Models (EHMM) or Embedded Bayesian networks for face recognition [94]. Experiments were carried out on the XM2VTS database, which resulted in error rates of 0.5% and 0.3% at various Gaussian noise levels.

### C. POST-MAPPING OR LATE FUSION

The post-mapping or late fusion based audio-visual fusion methods perform a data fusion on the results obtained individually from the classifiers of audio and visual domains. There are multiple ways of fusing the data in the late fusion approaches. The following sections describe all the late fusion methods used in audio-visual biometrics.

The popular methodologies considered combining the scores of audio and visual biometrics modalities using traditional mathematical rules. The sum and product rule are applied on face and voice recognition modalities built individually [1], [75]. Different face and voice recognition methods are examined, and the best performance of 87.5% subject acceptance rate was achieved by sum rule. Similarly, in [23], authors applied a weighted product approach for fusing scores from three visual classifiers and one acoustic classifier yielding an identification result of 98%. The Bayesian approach of decision fusion is another popular method used for the late fusion approach [41]. For speaker recognition, LPCs were used, and for face recognition, Gabor features are used. Experiments on the M2VTS database [102] displayed a success rate of 99.46% biometric authentication using the Bayesian supervisor. In [14], the scores obtained from face and voice modalities are efficiently fused to obtain a new optimal score that can be used for biometric recognition. Two individual (HMM for speech and EGM for face) recognition systems are used for fusion. Experiments on the M2VTS database gave a false alarm rate of 0.07% and a false rejection rate of 0% using linear SVM. The further works focused on text-dependent and text-independent speaker recognition [15]. Co-variance matrix on LPC feature vectors and arithmetic-harmony sphericity [18] measures are employed. Different fusion schemes are experimented on the XM2VTS database for person identity verification and observed the SVM-polynomial and the Bayesian classifiers displayed better results than other methods [15].

Hidden Markov models (HMMs) provided higher accuracy in performing speaker verification. In combination with GMMs, HMMs are used with the expectation-maximization algorithm in [35]. For face recognition, Eigenvectors are employed along with GMMs, and Bayes net is used to combine the confidence scores and the conditional probability distribution. The verification experiment yields good results for the combination of both modalities with a 99.5% success rate and 0.3% rejection rate per image, and a 100% verification rate per session, and a 99.2% recognition rate per image with 55.5% rejection rate per clip. The proposed text-independent module is robust to noise variations, and the Bayesian fusion method is a simple system that can select the most trustworthy images and audio clips from each session based on confidence scores.

Cepstral coefficients have proved to be performing well by representing speaker characteristics in automatic speaker verification. A late fusion based method with three strategies is used with a matrix of the codebook of vector quantized cepstral coefficients as speech features, and a synergic computer [50] for face recognition [50]. Vector Quantization is an efficient method to characterize the speaker's feature space and is used as a minimum distance classifier. The advantage of the synergetic computer is that it can build its own features; data reduction capability makes it suitable for face recognition. The three different fusion strategies defined determines the security risk.

Gaussian mixture models (GMMs) are used over HMMs for audio-based speaker recognition and Haar based face recognizer using regularised LDA (RLDA) and Recursive FLD (RFLD) as classifiers [89]. During fusion, the probability scores obtained from classifiers are combined to get the final audio-visual probability. Experiments were performed on AVIRES corpus [100]. The RFLD classifier performs better for face recognition on the AVIRES corpus when compared to RLDA classifier with an error rate of less than 15%. GMM based methods are also used along with a universal background model (UBM) for speaker recognition and LBP for face recognition [124]. The likelihood score from GMM-UBM and weighted distance metric on LBP are fused at the score-level. The fused method achieved an EER of 22.7% for males, 19.3% for females, and an average of 21.6%, which are far better than the EERs of individual cues. In [6], a novel Linear Regression Gaussian

Mixture Models along with Universal Background Model (LRC-GMM-UBM) is used for speaker recognition. For complementing the voice utterance, a Linear Regression-based Classifier (LRC) is used for face recognition. The scores from the two classifiers are normalized and fused using the sum rule. Experiments on AusTalk database [25] give an identification accuracy close to 100% and outperforms the fusion method as shown in [3]. In another kind of late fusion approach using the same recognition algorithms [3], a combination of a ranked list, which is a subtype of decision level fusion, is used.

Session variability modeling techniques built on the GMM baseline are examined in late fusion approach [73]. Inter Session Variability (ISV) [136], Joint Factor Analysis (JFA) and Total Variability (TV) [38] are the modelling methods used in this direction. DCT coefficients are used for face recognition to model Gaussian Mixture Models (GMM) and a pre-trained Universal Background Model (UBM). Session compensation techniques include Linear discriminant analysis [47] and WCCN normalisation [58]. After session compensation cosine similarity scoring [38] and Probabilistic Linear Discriminant Analysis (PLDA) [106] are used as scoring techniques. For fusing the face and voice modalities, Linear Logistic Regression (LLR) technique is used by combining the set of classifiers using the sum rule. Experiments were performed on the MOBIO database [84] for different protocols, and results indicate that ISV performs better compared to other compensation methods and the sum rule based fusion approach of all classifiers (GMM, ISV, TV) outperforms the ISV method in all protocols.

GMM-UBM based approach is used for both face [138] and speech authentication [90], [136]. GMM-UBM uses MAP adaptation, which is prone to changes in session variability and fewer enrollment data. These drawbacks were overcome using ISV modeling proposed in [73]. A weighted sum approach is used to fuse the scores from face and speech modalities. Initially, equal weights are assigned to the two classifiers, and an LLR method is used to learn the weights on the development set. Experiments on the MOBIO database [84] have resulted in an EER of 2.6% for males and 9.7% on female subjects. Similarly, Discrete Hidden Markov Models (DHMM) are used for both audio and visual domains [65]. Experiments were performed on VALID[3] audio-visual database and observed that the proposed fusion method is very adaptable for audio-visual biometric recognition method and can be used effectively in various authentication applications.

The deep learning approaches have paced into the biometrics research domain in recent years. In the early research on audio-visual biometrics using deep neural networks, two restricted Boltzmann machines are used to perform unsupervised training using local binary patterns for face, and GMM super vector for voice [4]. A squashing function called softmax layer is added on the top of DBM-DNN before they are fine-tuned discriminatively using a small set of labeled training data. The authors do not mention the amount of labeled data used for fine-tuning. The sum rule was used to fuse the scores from the outputs of DBM-DNN for each cue. Experiments on MOBIO and VidTIMIT datasets resulted in EERs of 0.66% and 0.84%, respectively. Hu *et al.* proposed a multimodal convolutional neural network (CNN) architecture to perform an audio-visual speaker naming [63]. A learned face feature extractor and audio feature extractor are combined with a unified multimodal classifier to recognize the speaker. Experiments on audio-visual data extracted from famous TV shows display an improved accuracy of 90.5% over 80.8% from previous methods. Authors have also emphasized that even without face tracking, facial landmark localization, or subtitle/transcript, the proposed method achieved an accuracy of 82.9%. The latest method on a late fusion technique used similarity matrix of MFCC voice features and SVM face scores from personal devices [87]. The Lagrangian multiplier of SVM is used for fusing the scores, and the accuracy of 73.8% is obtained.

Quality assessment based score-level fusion was performed by Antipov *et al.* [11] for Audio-Visual speaker verification. For face recognition, four face embeddings, namely ResNet-50, PyramidNet, ArcFace-50, ArcFace-100, are aggregated using a Transformer aggregation model. For speaker recognition, six variants of X-vector based methods are fused using Cllr-logistic regression. The audio-visual speaker verification is performed by performing a score-level fusion of verification scores and quality of enrolment and test sample in face and speech modalities. Experiments were performed on different types of quality fusion methods compared to the baseline of sum-rule based fusion. Results indicate that using all quality estimates improve speaker verification performance.

An overview table 2 summarises the AV person biometric systems discussed in this survey paper.

## V. AUDIO-VISUAL BIOMETRIC DATABASES

Widely variety of audio-visual databases were created by capturing talking persons' videos focusing on face and voice modalities. Multimodal databases include modalities like a fingerprint, face, iris, and biometric voice data. However, our study focuses on audio-visual databases, which include only face and voice modalities. This section presents a detailed study on both publicly available audio-visual biometric databases. A comparison of databases with each other can be found in Table 4. Some databases, like DAVID [32] is mentioned in other works, but no published work is found. Other databases like DaFEx [13] contain audio-visual data but not recorded for the application of biometrics. Therefore, they are not discussed in this report.

*AMP/CMU Dataset:* The advance multimedia processing (AMP) lab of Carnegie Melon University (CMU) has created an audio-visual speech dataset that contains ten subjects (seven male, three female).[4] Each subject speaks 78 isolated words, and a digital camcorder with a tie-clip microphone is used to record [133]. The sound file and extracted lip parameters are available to the public domain, and video data is available upon request.

---

[3]The VALID database: http://ee.ucd.ie/validdb/

**TABLE 2.** Overview table showing features used, classifier fusion method, database, number of subjects, performance achieved, recognition type starting from the year 1995 to 2018. *TD:text-dependent, *TI:text-independent, *SEP:Standard Evaluation Protocol, *Dev:Development, *E:Evaluation, *F:Female, *M:Male.

| Authors | Features used | | Classifier | AV fusion method | Database used | No. of subjects & sessions | Performance achieved | Recognition type |
|---|---|---|---|---|---|---|---|---|
| | Audio | Visual | | | | | | |
| Brunelli et al. [23] | MFCC | Animation | VQ Comparision at pixel level | Weighted product | Self accquired with CCD camera | 89 3 sessions | Recognition rate: 98% | Identification |
| Kittler et al. [1] | LPC | Gabor | HMM Gabor matching grid | Sum rule | M2VTS | 37 16 users 21 imposters | Acceptance rate: 87.5% | Verification |
| Duc et al. [41] | LPC | Gabor | HMM EGM | Bayesian fusion | M2VTS | 5328 Both client and imposter | Success rate: 99.46% | Verification |
| Dieckmann et al. [40] | Fourier transform | Optical flow | MELT algorithm | Sensor fusion | Self accquired with CCD camera | 66 15 clients 26 common 25 test | Identification rate: 93% Verification rate: 99.8% | Identification Verification |
| Yacoub et al. [14] | LPC | Gabor filter | HMM EGM | SVM | XM2VTS | 295 200 clients 25 evaluation imposters 70 test imposters | EER: 0.58% | Verification |
| Choudhury et al. [35] | MFCC | Optical flow | HMM Eigen vectors | Bayes net | Automated teller machine | 26 | Verification rate:99.5% Recognition rate: 99.2% | Verification Identification |
| Robert et al. [50] | Cepstral coefficients | FFT | VQ Synergetic computer | Sensor fusion | Self testing | 150 | FAR: < 1% | Verification |
| Nefian et al. [94] | MFCC | Haar-like | CHMM EHMM | Score level | XM2VTS | 348 files (training) 320 files (testing) | EER: 0.5% | Identification |
| Bengio et al. [16] | MFCC | ASM | AHMM | Midst mapping | M2VTS | 2 sessions (client model) 3 sessions (testing) | HTER: 15% | Verification |
| Isaac et al. [46] | MFCC | Optical flow | GMM | Feature level | XM2VTS | 200 (training) 25 (test, evaluation imposters) | EER: Evaluation: 1% Test:2% | Verification |
| Shah et al. [123] | MFCC | DCT | GMM | Feature level | VidTIMIT | 43 35 clients 8 imposters | FAR: 1% client FAR: 0% Imposters | Verification |
| Micheloni et al. [89] | MFCC | Haar-like | GMM RFLD | Score level | AVIRES | 6 | Classification error: 15% | Verification |
| Sugiarta et al. [143] | DTCWPT | DTCWT | PCA | Feature level | VidTIMIT | Session1,2 (training) session-3 (testing) | Identification: rate: 90% (TD) 93.7% (TI) | Identification |
| Jiang et al. [69] | MFCC | Haar-like | Laplacian Eigenmap | Feature level | Open web tv | 10 8 (training) 2 (testing) | EER: 35% EER | Verification |
| Shen et al. [124] | MFCC | LBP | GMM-UBM LBPH | Score level | MOBIO | 160 session-1 (enrolment) session 2-6 (testing) | EER: M: 22.7%; F: 19.3% Average: 21% | Verification |
| Chenxi et al. [144] | MFCC | Pyramidal Gabor filter | PNN | Feature level | Virtual subjects | 40 | Recognition rate: 100% | Identification |
| Motlicek et al. [90] | MFCC | DCT | GMM | LLR score level | MOBIO | Protocols: mobile-0, mobile-1, laptop-1, laptop-mobile-1 | EER: Dev: M:1.2%; F:2.3% Test: M:2.6%; F:9.7% | Verification |

**TABLE 2.** *(Continued.)* Overview table showing features used, classifier fusion method, database, number of subjects, performance achieved, recognition type starting from the year 1995 to 2018. *TD:text-dependent, *TI:text-independent, *SEP:Standard Evaluation Protocol,*Dev:Development, *E:Evaluation, *F:Female, *M:Male.

| | | | | | | | | |
|---|---|---|---|---|---|---|---|---|
| Xuran *et al.* [145] | MFCC | LBP | MSDA | Feature level | MOBIO | session-1,2,3 training session-8,9,10 testing | Recognition rates: session-1: 90.6% session-2: 96.7% session-3: 97.4% | Verification |
| Alam *et al.* [3] | MFCC | Haar-like | LRC-GMM-UBM LRC | Ranked list | Austalk | 88 | Identification accuracy: 31.3% | Identification |
| Khoury *et al.* [73] | MFCC | DCT-mod2 | GMM+ISV+TV | LLR score level | MOBIO | Protocols: mobile-0 (m0), mobile-1 (m1), laptop-1 (l1), laptop-mobile-1 (lm1) | EER: (Dev; Eval) m0: F: (1.43%; 6.30%) M: (0.92%; 1.89) m1: F: (1.64%; 6.32%) M: (0.75%; 2.06%) l1: F: (2.91%; 6.83%) M: (1.82%; 3.37%) lm1: F: (1.11%;6.32%) M: (0.64%;1.77%) | Verification |
| Alam *et al.* [6] | MFCC | Haar-like | LRC-GMM-UBM LRC | Sum rule | Austalk | 88 | Accuracy: 100% | Identification |
| Tresadren *et al.* [132] | MFCC | LBP | Boosted slice classiifier | QDA | MoBio | - | EER: 0.5% | Verification |
| Shi *et al.* [125] | MFCC | LBP | DTW LBPH | Sum rule | Self accquired | 11 | Authentication Error: 0% | Verification |
| Islam *et al.* [65] | MFCC | ASM | HMM | BPN | Self accquired | 11 | Authentication Error: 0% | Verification |
| Alam *et al.* [4] | MFCC | LBP | DBM-DNN | Feature level | VidTIMIT MOBIO | Protocol 1: train:session-1+2 test: session-3 Protocol:2 train: session-1+3 test:session-2 SEP | EER: Protocol1: 0.66% Protocol: 0.84% | Verification Identification |
| Primorac *et al.* [105] | MFCC | Haar-like | Joint sparse classifier | Feature level | MOBIO | SEP | Recognition rate: 0.942 | Verification |
| Alam *et al.* [5] | MFCC | LBP | jDBM | Feature level | MOBIO | SEP | Identification: rate: 99.70 (TD) 96.30 (TI) | Identification |
| Memon *et al.* [87] | MFCC | SVM scores | Similarity matrix | Feature level | Self accquired | 15 | Accuracy :73.8% | Verification |
| Gofman *et al.* [53] | MFCC | HOG | SVM | Feature level | CSUF-SG5 | 27 | EER: 20.59% | Verification |
| Antipov *et al.* [11] | Four CNN methods | Six X-vector variants | Logistic Regression | Score level | NIST SRE19 | M/F: 15/32 (Dev) M/F: 47/102 (Test) | EER: Dev: 2.78% Test: 0.6% | Verification |

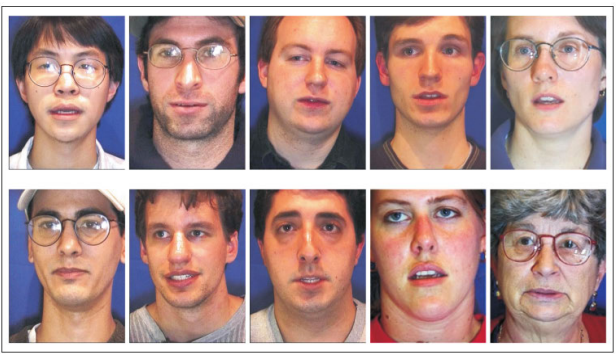

**FIGURE 6.** Example AMP/CMU dataset images [133].

Aleksic *et al.* [7] used 13 MFCC coefficients with first and second-order derivatives, audio features, and a visual shape-based feature vector of ten Facial animation parameters (FAPs) to develop an AV speaker recognition system. Fusion integration approach is employed with single-stream HMMs and speaker verification and identification experiments performed on the AMP/CMU dataset. The results of audio-only and audio-visual speaker recognition at different signal-to-noise ratios (SNRs) are presented in Table 3.

*The BANCA Database:* Biometrics Access Control for Networked and E-Commerce Applications (BANCA)[5] [12] is one of the earliest audio-visual datasets used for E-Commerce applications. Two modalities of face and voice were captured in four European languages with both high and low-quality microphones and cameras. Throughout capturing, three different scenarios, controlled, degraded, and adverse, are included in 12 different sessions for three months. The total number of subjects was 208, with an equal number of men and women. Figure 7 shows the example images of database captured in three different scenarios. The database is benchmarked with a weighted sum rule score-level fusion technique. The features used are DCT-mod2 for face and MFCCs for voice. The GMM models are used to perform face and voice classification, and audio-visual speaker verification obtained an equal error rate of 3.47% without impostors.

---

[4]The AMP/CMU dataset: http://amp.ece.cmu.edu/

[5]The BANCA database: http://www.ee.surrey.ac.uk/CVSSP/banca/

**TABLE 3.** Comparison of audio-only (AU) and audio-visual (AV) speaker recognition performance proposed in [7].

| SNR | Identification Error (%) | | Verification Error (%) | |
| --- | --- | --- | --- | --- |
| | AU | AV | AU | AV |
| 30 | 5.13 | 5.13 | 2.56 | 1.71 |
| 20 | 19.51 | 7.69 | 3.99 | 2.28 |
| 10 | 38.03 | 10.26 | 4.99 | 2.71 |
| 0 | 53.10 | 12.82 | 8.26 | 3.13 |

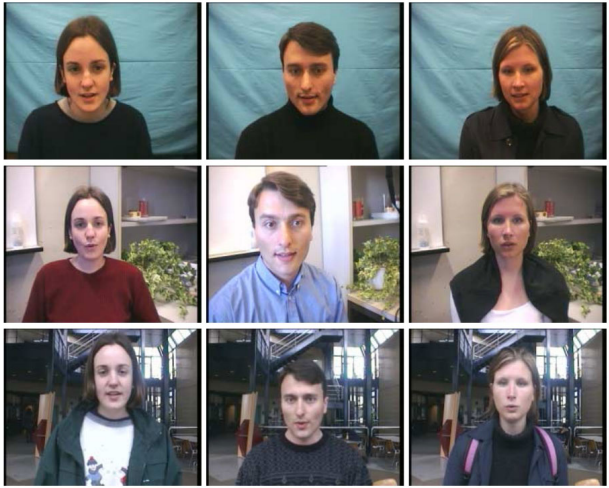

**FIGURE 7.** Example BANCA database images Up: controlled, middle: degraded and Down: adverse scenarios [12].

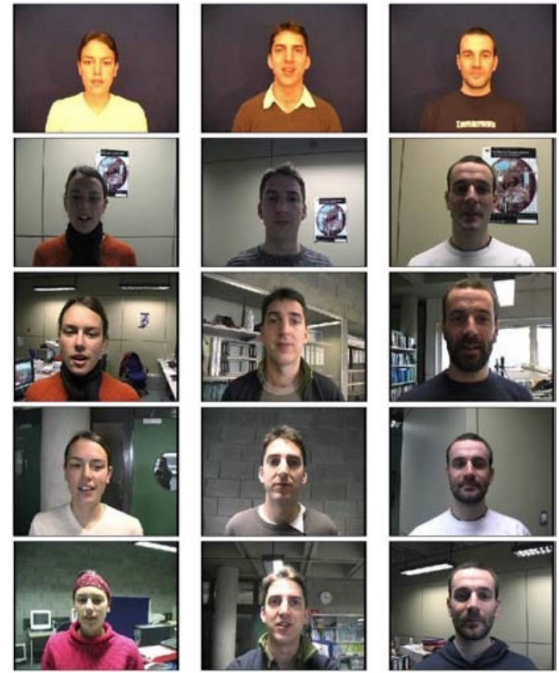

**FIGURE 8.** Three VALID database subject images from each of the five sessions [49].

*The VALID Database:* The aim of the VALID database is to provide robust audio, face, and multimodal person recognition systems. Therefore, the VALID database was acquired in a realistic audio-visual noisy office scenario with no control over lights or acoustics. This database is captured in five sessions with 106 subjects for a period of one month. The performance degradation of the uncontrolled VALID database is observed in comparison to that of the controlled XM2VTS database [49]. The VALID database is publicly available to the research community through the websites.[6] Figure 8 shows the example images from the VALID database captured in five different sessions.

[6]The VALID database: http://ee.ucd.ie/validdb/

The audio-visual experiments are performed on the VALID database to address noise problems in a single modal speaker identification [65]. A new score fusion approach is proposed using a back-propagation learning feed-forward neural network (BPN). The verification results from appearance-shape based facial features and MFCC based audio features are combined using BPN score fusion, and a speaker identification of 98.67% is achieved at SNR of 30dB.

*The M2VTS Database:* The MultiModal Verification for Teleservices and Security applications database has developed with the primary goal of issuing access to secure regions using audio-visual person verification [103]. Five shots were taken for each of the 37 subjects, with an interval of one week between each shot. The camera used for shooting the face images is a Hi8 video camera. D1 digital recorder is utilized for recording and editing the voice. The voice recordings are captured by speakers uttering the numbers from 0 to 9 in their native language (mostly French). The M2VTS database is available to any non-commercial user on request to the European Language Resource Agency.

Multimodal data fusion using support vector machines (SVM) method used the M2VTS database to perform audio-visual person identification [14]. The experiments display a dominance of SVM performance over Bayesian conciliation, speech only, and face only experts. This approach's face features are based on Elastic Graph Matching (EGM), and speech features are Linear Predictive Coefficients-Cepstrum (LPC-C). The total error rate (TE), which is a sum of false acceptance (FA) rate and false rejection (FR) rate, is computed, and Linear-SVM gave the least TE of 0.07%.

*The XM2VTS Database:* An extension to the M2VTS database with more subjects and latest devices, XM2VTS (extended M2VTS) is focused on a large multimodal database with high-quality samples [88]. This database contains four recordings of 295 subjects taken during four months. Each recording contains a speaking headshot and a rotating headshot. The data comprises high-quality color images, 32 kHz 16-bit sound files, video sequences, and a 3D Model. XM2VTS database is used in many research works for AV speaker verification. The database is made publicly available at cost price only.[7]

Different fusion approaches are experimented on XM2VTS database for person identity verification [15]. The elastic graph matching (EGM) based face features are computed, and two voice features, namely sphericity, and hidden Markov models (HMM), are used for six different fusion classifiers (SVM-polynomial, SVM-Gaussian, C4.5, Multilayer perceptron, Fisher linear discriminant, Bayesian classifier). It is observed that the bayesian fusion method with different combinations of face and voice features (with text-dependent and text-independent scenarios).

A coupled HMM (CHMM) is used as a classifier as audio-visual speech modeling for speaker identification. 2D discrete cosine transform (2D DCT) coefficients are used as facial features and MFCCs as acoustic features. The visual speech features are computed from the mouth region through a cascade algorithm. Finally, the audio features and visual features are combined using a CHMM. A two-stage recognition process is performed by computing face likelihood using embedded HMM and audio-visual speech likelihood using CHMM separately. The face-audio-visual speaker identification system is created by combining face and audio-visual speech likelihoods and has achieved an error rate of 0.3%.

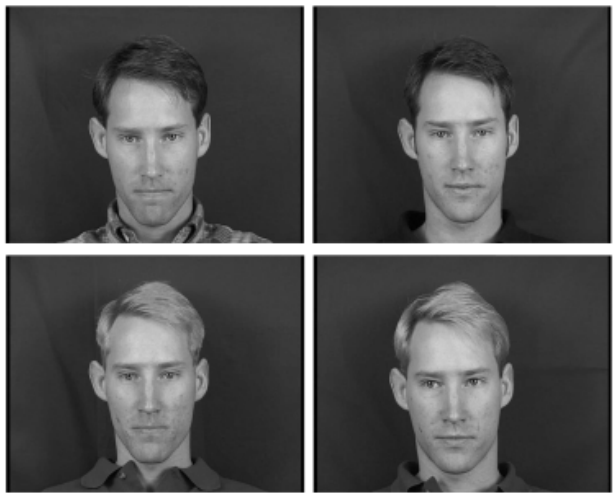

**FIGURE 9.** Front profile shots of a subject from four sessions of XM2VTS database [88].

*VidTIMIT Database:* Video recordings of people reading sentences from Texas Instruments and Massachusetts Institute of Technology (TIMIT) corpus (VidTIMID)[8] is a publicly available dataset for research purposes [118]. The dataset is captured in 3 sessions with a mean delay of 6-7 days. Each person reads ten sentences that include alpha-numerical utterances, with the first two sentences the same for all subjects. Along with the sentences, a head rotation sequence is recorded for each person in each session [117].

VidTimit dataset is used in audio-visual person recognition using deep neural network [4]. The local binary patterns (LBPs) as visual features and gaussian mixture models (GMMs) built on MFCCs have used speech features. The Deep Boltzmann Machines based deep neural network model (DBM-DNN) is used to compute scores from extracted features fused using a sum rule. The audio-visual based speaker recognition has improved the performance over the single modal recognition with an EER of 0.84%.

Liveness detection is another prominent area where VidTIMIT is employed. Gaussian mixture models [29], [30], Cross-modal fusion [28] and delay estimation methods [147] are experiments on VidTIMIT dataset to perform replay attack detection using audio-visual complimentary data.

*BioSecure Database:* BioSecure[9] is another popular multimodal database contains different biometric modalities and can be used as a audio-visual dataset [95]. The database consists of data from 600 subjects recorded in three different scenarios. The sample images from the database are shown in Figure 10.

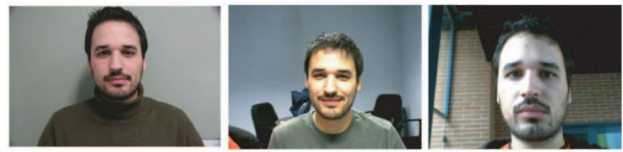

**FIGURE 10.** Face samples acquired in BioSecure database in three different scenarios. Left: indoor digital camera (from DS2), Middle: Webcam (from DS2), and Right: outdoor Webcam (from DS3) [95].

*AVICAR*[10]: AVICAR is a public audio-visual database captured in a car environment through multiple sensors consisting of eight microphones and four video cameras [77]. The speech data consists of isolated digits, isolated letters, phone numbers, and sentences in English with varying noise.

*MOBIO Database*[11]: The MOBIO database [84] is a bi-modal (audio and video) data collected from 152 people with 100 males and 52 females. It is captured at six different sites from five different countries in 2 phases (6 sessions in each phase). This database's important feature is that it was recorded using two mobile devices: a mobile phone (NOKIA N93i) and a laptop computer (2008 MacBook).

---

[7] The XM2VTS database: http://www.ee.surrey.ac.uk/CVSSP/xm2vtsdb/
[8] The VidTIMTI dataset: http://conradsanderson.id.au/vidtimit/
[9] BioSecure: https://biosecure.wp.tem-tsp.eu/biosecure-database/
[10] AVICAR Database:
[11] The MOBIO database:

MOBIO dataset helped in the study of person identification in a mobile phone environment. Session variability modeling is used to perform bi-modal authentication using the MOBIO database. Inter-session variance is exploited to compensate for the drawbacks of GMM-UBM based methods, and a weighted sum rule based fusion [90]. Using face features like DCT with GMM modeling and sum rule based fusion displayed an improvement in person authentication [73]. Further, deep learning methods have also used the MOBIO database for experiments on AV person recognition. DBM-DNN [4] and jDBM [5] methods are utilised on MOBIO dataset and displayed improved person identification.

*MobBIO Database:* The MobBIO database consists of face, iris and voice biometrics from 105 volunteers (29% females and 71% males) [122]. The data capturing process took place in 2 different lights. The device used is the rear camera of the Asus Transformer Pad TF 300T for capturing 16 faces and 16 iris images. Each volunteer was asked to read 16 sentences in Portuguese for voice biometrics.

*Hu et al. Dataset:* A new audio-visual dataset was recently captured by Hu *et al.* [63], which is used in developing a deep learning-based feature fusion. The database is acquired from three hours of videos of nine episodes from two popular television shows with annotated subjects. Face and audio of six people from ''Friends'' and five from ''The Big Bang Theory'' are annotated and provided in this dataset. Two initial experiments are used, namely, face only recognition and identifying non-match face-audio pairs to improve audio-visual recognition performance (speaker naming). In the speaker naming process, a neural network approach is used to identify the speaker in each frame using a matched face-audio pair. This method has achieved an accuracy of 90.5%.

*AusTalk database:* Australian Speech Corpus (AusTalk)[12] provides the data of people reading predefined set of sentences in English [25]. The database is a part of the Big Australian Speech Corpus project consisting of speech from 1000 geographically and socially diverse speakers and recorded using a uniform and automated protocol with standardized hardware and software. A linear-regression based classifier it used for audio-visual person identification on AusTalk database achieving 100% accuracy [6].

*SWAN Database:* The Smartphone Multimodal Biometric database was collected to meet the real-life scenarios such as mobile banking [109]. The database was captured in six different sessions and four locations using iPhone 6s and iPad Pro cameras. The database consists of audio-visual data of 150 subjects with English as a common language and Norwegian, French, and Hindi as secondary languages. Figure 11 shows the sample images of subjects from six sessions.

*NIST SRE19 AV Database:* This database contains the videos from the VAST portion of the SRE18 development set. This database's development set is publicly available for the 2019 NIST Audio-Visual Speaker Recognition

[12]AusTalk database: https://austalk.edu.au/

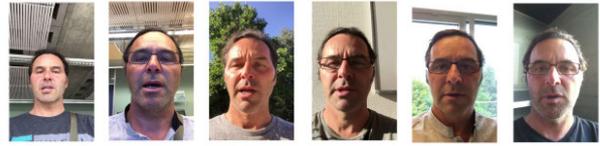

**FIGURE 11.** Talking face samples from SWAN database one frame from each session [109].

Evaluation [116]. The videos are in interview-style and are similar to the VoxCeleb database. The videos are incredibly diverse in quality and acoustics because they are recorded mostly using personal handheld devices like smartphones. The videos contain manually diarization labels as the videos may contain multiple speakers.

The 2019 NIST Audio-Visual SRE challenge has releases results of the top-performing submissions in AV recognition. The approach used by the top-performing method is unknown. However, the results show that combining face and speaker recognition systems have displayed an increase of 85% of minimum detection cost compare to face or speaker recognition system alone. The EER for AV speaker recognition achieved by the top-performing team is 0.44%.

## VI. PRESENTATION ATTACK DETECTION (PAD) ALGORITHMS

Audio-visual biometrics are vulnerable to various artifacts that can be generated with less cost. So it is necessary to identify and mitigate these attacks to enhance both the security and reliability of AV recognition systems. This section presents a thorough review of existing presentation attack detection (PAD) algorithms against replay attacks and forgery attacks for AV biometrics. Although there are many attack detection algorithms in single biometrics, like ASVSpoof [130], we have only included the audio-visual PAD methods in this section. The main intention is to take advantage of bimodal biometric characteristics to optimize attack detection algorithms.

### A. AUDIO-VISUAL FEATURES USED FOR LIVENESS DETECTION

Many works in AV biometrics suggested the liveness detection technique, which acts as a guard for possible replay attacks against the audio-visual recognition system. The fig 12 shows different features used for PAD in audio-visual biometrics.

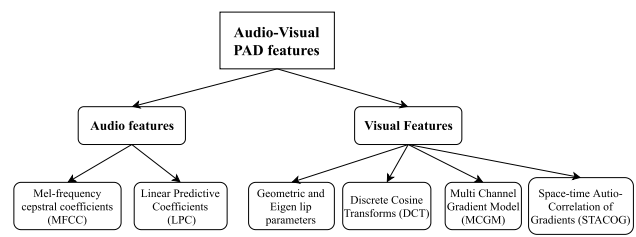

**FIGURE 12.** Different audio-visual features used in PAD.

**TABLE 4.** Details of Audio-visual Biometric Verification Databases.

| Dataset | Year | Devices | No. of subjects | Best Performing Algorithm | Accuracy |
|---|---|---|---|---|---|
| AMP/CMU [133] | 2001 | Digital Camcorder, tie-clip microphone | 10 (7 M, 3 F) | MFCC + FAPs [7] | EER = 3.13% |
| BANCA [12] | 2003 | Webcam and Digital Camera | 208 (104 M, 104 F) | DCT-mod2 + GMM [72] | EER = 3.47% |
| VALID [49] | 2005 | Canon 3CCD XM1 PAL | 106 (77 M, 29 F) | BPN score fusion [66] | Accuracy = 98.67% |
| M2VTS [103] | 2005 | Hi8 camera, D1 digital recorder | 37 | EGM + LPC [14] | Total error rate = 0.07% |
| XM2VTS [88] | 2005 | Sony VX1000E, DHR1000UX | 295 | CHMM [95] | Error rate = 0.3% |
| VidTIMIT [118] | 2009 | Digital video camera | 43 (24 M, 19 F) | FAPs + MFCC [7] | EER = 1.71% |
| BioSecure [95] | 2010 | Samsung Q1, Philips SP900NC HP iPAQ hx2790 Webcam, PDA | DS1: 971 DS2: 667 DS3: 713 | - | - |
| AVICAR [77] | 2010 | Multiple, sensors | 100 (50 M, 50 F) | - | - |
| MOBIO [84] | 2012 | Nokia N93i Mac-book | 152 | jDBM [5] | Accuracy = 99.7% |
| MobBIO [122] | 2014 | Asus Transformer Pad TF 300T | 105 | - | - |
| Hu et al. [63] | 2015 | - | 11 | Deep Multimodal Speaker Naming [64] | Accuracy = 90.5% |
| AusTalk [25] | 2016 | Black Box | 88 | LRC-GMM-UBM [6] | Accuracy = 100% |
| SWAN database [109] | 2019 | iPhone 6 iPad Pro | 88 | FaceNet+DRN [110] | EER = 3.1% |
| NIST SRE19 AV database [116] | 2019 | Multiple devices | 15, 37 (M, F) | Anonymous [117] | EER = 0.44% |

### 1) MEL-FREQUENCY CEPSTRAL COEFFICIENTS

MFCCs are popular feature vectors used for both speaker recognition and liveness detection. There are different visual features used alongside MFCCs to perform reliable liveness detection. They include Geometric lip parameters and Eigen lips, Discrete cosine transforms (DCTs), Multi-Channel Gradient Model (MCGM), and Space-Time Auto Correlation of Gradients (STACOG).

Geometric lip parameters used for AV liveness verification are heights and widths of inner, outer, lower, and upper lip regions [29]. Also, Eigen lip representation is used for complementing the MFCCs parameters in this method. The advantage of this method is that the alternate color spaces in an image are exploited compared to deformed images, and it can be extended to detect and extract multiple faces and their features with different backgrounds. In further works of Chetty et al., a multi-level liveness verification is proposed by exploiting correlation between the cues using MFCCs, Eigen lip, 3D shape and textures features of the face with the help of different fusion techniques [30].

DCT coefficients on lip regions are used as visual features complementing MFCCs for liveness detection in [114]. Further, a client dependent synchronous measure is introduced using the Voila-Jones algorithm [134] for detecting face region and extracting first order DCTs [22]. DCT-mod2 [119] coefficients are another face image representations computed on normalized faces for robust audio-visual biometric systems against forgery attacks [71]. Another approach used DCT coefficients extracted from mouth region with Least Residual Error Energy (LREE) algorithm [34] and MFCCs as audio features [147].

Multi-Channel Gradient Model (MCGM) algorithm is a neurological and psychological algorithm used to build artificial vision systems [86]. MCGM uses gradient methods, which computes the motion as a ratio of partial derivatives of input image brightness concerning space and time. MCGM is used in cross-modal fusion method for biometric liveness verification using kernel Canonical Correlation Analysis (kCCA) [28]. This method aims at extracting the non-linear correlation between audio-lip articulators and lip motion features from MCGM. The audio-visual cues are mutually exclusive; hence a statistical technique called Independent Component Analysis (ICA) is used. The advantages of cross-modal fusion are that it exploits mutually independent components from face and voice cues in Spatio-temporal couplings and extracts correlated information.

Space-Time Auto Correlation of Gradients (STACOG) is a motion feature extraction method that uses space-time gradients of three-dimensional moving objects in a video. STACOGs are used for measuring audio-visual synchrony to discriminate live and biometric artifact samples [20]. STACOG utilizes auto-correlation to exploit the local-relationship, such as co-occurrence among space-time gradients. STACOG also exploits local geometric characteristics and possesses shift-invariance, which is a useful property for biometric recognition.

### 2) LINEAR PREDICTOR COEFFICIENTS

Linear prediction model predicts the next point as a linear combination of previous values (see section III-A4). In complementing LPCs as audio features, geometric lip parameters are extracted from the jumping snake algorithm [45], which achieves lip segmentation. These parameters are used with a co-inertia analysis for the liveness test in audio-visual biometrics [44]. Three video features are extracted from the width, height, and area of the mouth region. The audio-visual features used here exhibit a tight link between the lip contour and speech produced to detect liveness effectively.

### B. LIVENESS DETECTION METHODS FOR REPLAY ATTACKS

This section discusses the liveness detection methods used for identifying replay attacks. As the audio-visual biometrics contain complementary information, research works used different approaches to make use of both modalities in effectively detecting replay attacks.

Gaussian Mixture Models (GMMs) comprise audio-visual feature vectors trained for each client and used as a countermeasure for replay attacks [29]. For testing clients, a log-likelihood is computed against the client model. Three experiments were conducted for live and four replayed recordings, and performance is calculated. The results of liveness detection show that using Eigen lip projections, lip contours with MFCCs gives an EER for less than 1% for all cases.

The Co-Inertia (CoIA) and Canonical Correlation Analysis (CANCOR) are statistical methods used to measure the relationship between two multidimensional data. They are computed using Pearson correlation projections are used for liveness detection in AV biometrics [44]. Ordinary correlation analysis is dependent on the coordinate system in which the variables are described, where CANCOR and CoIA focus on finding the best coordinate system, which is optimal for correlation analysis. CoIA method has numerical stability and does not suffer from collinearity. Experiments were conducted on the XM2VTS database with two kinds of replay attacks created with the same and different sentences uttered. The result shows that CoIA method gives EER values of 14.5% and 12.5%, where CANCOR method shows 23.5% and 22.5% on replay attack 1 and replay attack 2, respectively.

Multi-level liveness verification is proposed using three different fusion techniques, namely Bi-modal feature fusion (BMF), Cross-Modal Fusion (CMF), and 3D multimodal fusion (3MF) [30]. Experiments were performed on Vid-TIMIT, UCBN, and AVOZES. A 10-mixture Gaussian mixture model for each client and each fusion approach is trained by constructing a gender-specific UBM and then adapting each UBM with MAP adaptation. While testing, the client's live recordings were evaluated against the client's model by calculating the log-likelihood of audio-visual vectors. Three types of replay attacks were created for testing:

**TABLE 5.** Performance of liveness verification techniques proposed in [30] (EER%).

| Approach | Photo replay | Video replay | Synthetic replay |
|---|---|---|---|
| BMF | 2.4% | 6.54% | 9.23% |
| CMF | 0.29% | 2.25% | 3.96% |
| 3MF | 0.0155% | 0.611% | 1.18% |

photo replay attack, video replay attack, and synthetic replay attack. It is observed that photo replay attacks are easy to detect compared to the other two attacks and the 3MF fusion method is more robust than other methods. The equal error rates of the three proposed techniques on three types of replay attacks is shown in Table 5.

The synchronous information between audio and visual cues is an advantageous detail that can be used in liveness detection. By measuring the asynchrony, a presentation attack can be detected. The degree of synchrony between lips and voice in a video sequence is used for liveness detection in [114]. The methods used are co-inertia analysis (CoIA) and coupled hidden Markov models (CHMM). Three different methods of CoIA were proposed, namely world training model, self-training method, and piece-wise self-training method. A CHMM is a collection of HMM which uses the Baum-Welch algorithm for training. Further, the Viterbi algorithm calculates the states for every stream and the frame likelihoods. CoIA and CHMM methods are fused using Bayesian fusion [57]. Experiments were performed on the BANCA database [12] using two protocols: controlled and pooled. Only recordings from controlled conditions are used from the BANCA database's world model in the controlled protocol. In the Pooled protocol, three conditions, such as controlled, adverse, and degraded, are used. The sum rule fusion of CoIA and CHMM methods in controlled protocol resulted in lower error rates than individual methods. The CHMM method displayed the lowest error rates in detecting replay attacks in the pooled protocol.

In similar fashion, a client dependent synchrony measure is introduced to thwart the deliberate impostor attacks [22]. For the extracted acoustic and visual features, CoIA is applied, which maximizes the covariance of AV features in the enrollment phase. While testing for the AV features, a correlation measure based on CoIA is computed. This method produces a weighted error rate (WER) of 7.7% for random impostors and 6.9% for deliberate impostors. Since WER for Random impostor attacks is on the higher side, three other fusion strategies are proposed. The first fusion strategy is the weighted sum of scores of speech, face verification, and synchrony which makes the method sensitive to deliberate impostor attacks. The second fusion strategy aims to reduce the first strategy score with a low synchrony verification score, making it robust to random impostor attacks. The third fusion strategy is an adaptive weighted sum of normalized scores. More weight is given to the synchrony verification module if the synchrony score is the least, and weight is decreased if the synchrony score is high. The three fusion strategies proposed makes the system robust to deliberate imposters.

The cross-modal fusion based on Bayesian Fusion is adapted for Liveness detection in [28]. The audio module, PCA Eigen lip module, kCCA module, and ICA module are summed up as logarithmic class conditional probability and fused under the reliability weighted summation (RWS) rule. Experiments were performed on VidTIMIT and DaFEx databases [81] using 10-mixture GMMs and log-likelihood scores. Two types of attacks were tested: static replay attack displaying the still photo and dynamic attack where faces are synthesized from still photos. The single-mode features such as MFCC, PCA, Eigen lips produce higher errors in detecting the attacks. However, when MFCC, PCA, Eigen lips, kCCA, and ICA were fused, the method produced a promising performance.

A delay estimation method is a process of shifting audio features positively and negatively to check for the liveness in audio-visual samples [147]. Experiments are conducted on the VidTIMIT database using three types of inconsistent data and time delay based scoring. Three types of attacks include audio-video from the same subject but a different sentence, audio-video from different subject and sentence, and finally, audio-video from a different subject but the same sentence. When experimented on Co-Inertia Analysis (CoIA) and Canonical Correlation Analysis (CCA), upon adding the delay estimation method, the performance of liveness detection has improved.

The Space-Time Auto-Correlation of Gradients (STACOG) is used for measuring the audio-visual synchrony in [20]. Two cross-modality mapping approaches are used to estimate synchrony: Partial Least Square Analysis (PLS) and Canonical Correlation Analysis (CCA). PLS method [112] models the input and output onto a low-dimensional subspace. The projections are chosen such that covariance between input and output scores are maximized. CCA is a statistical method used to measure the relationship between two multidimensional data. The correlation between the audio-visual feature vectors is obtained from either CCA or PLS, as described in [22]. Experiments were performed on the BANCA and XM2VTS database with four different kinds of replay attacks. The proposed method with STACOG for visual speech features and CCA for acoustic features produced promising results proving the advantage of visual speech joint features.

### C. FORGERY ATTACKS IN AV BIOMETRICS
Forgery attacks are performed by digitally transforming both voice and face cues. There are no liveness detection approaches in the literature; however, in this section, we discuss the impact of forgery attacks on AV biometrics.

The robustness of the audio-visual biometric systems against forgery attacks is examined in [71]. The forgery attacks are created using a mixture-structured bias voice transformation technique called MixTrans. This method allows the transformed signal to be estimated and reconstructed in the temporal domain. Once the transformation is defined as bias, it makes the source client vectors resemble a target client, and a maximum likelihood criterion is used to estimate the transformation parameters. Finally, a synthesis step is used to replace the source characteristics with those of the target speaker. Face transformation is performed by a MPEG-4 face animation based approach using a thin-plate spline warping. Experiments were performed on the BANCA database consisting of 7 distinct training and testing configurations. The proposed forgery attack shows an increase in EER of the AV identity verification system when compared to the AV system with no attacks. For two groups of the dataset, EER has increased from 4.22% to 11% and from 3.47% to 16.1% with and without forgery attacks, respectively. The vulnerability of AV biometrics to forgery attacks is high, and there is a strong requirement for forgery attack detection methods in AV systems.

Table 6 summarises the different AV features, PAD algorithms, different databases, EERs achieved in the AV recognition system.

## VII. CHALLENGES AND OPEN QUESTIONS
Audio-visual biometrics has gained intensive research efforts in terms of developing novel recognition systems and PAD algorithms to thwart the artifacts. Despite all these, there are several challenges and open questions in AV biometric methods. In this section, we discuss some well-known challenges and open research questions.

### A. DATABASES AND EVALUATION
The publicly available databases [12], [49], [84], [88], [95], [103], [118], [122] are usually recorded using limited number of devices and sessions. The lack of variance in biometric data challenges the development of robust AV biometric algorithms. This gives rise to the problem of generalization of a biometric algorithm. For example, in smartphone biometrics, it is necessary to have an AV-based recognition algorithm that is adaptable to changes in the recording device. Therefore the databases recorded using multiple devices and in different sessions help in improving the robustness of recognition systems. Further, the mismatches arise when the enrolled sample is from one type of device and tested with another. The change in devices also introduces the problem of cross-device recognition errors. There is a requirement of AV databases that includes different types of biometric dependencies like device, lighting, background noise.

#### 1) PRESENTATION ATTACK DATABASE FOR AV BIOMETRICS
Available databases are also limited in terms of various kinds of presentation attacks. Moreover, data for many attacks specified in the literature are not publicly available. There are new kinds of attacks being generated that pose a huge threat to biometric systems in both audio [130] and face [108]. For example, voice impersonation has shown to be causing a considerable vulnerability to automatic speaker recognition [83]. However, the databases or protocols to create such attacks are not publicly available. Therefore, this is a hindrance for the researchers to develop robust PAD algorithms. AV biometric systems can also be attacked in either of the audio, visual, or combined audio-visual (bi-modal) domains. Depending on the authentication algorithm, a good bi-modal attack can pose a severe threat to the AV system. So there is a need to create a high-quality presentation attacks database and make it available for research.

**TABLE 6.** Table showing summary of different features, methods for liveness detection, databases used and EERs achieved. (Attack type: Replay attack).

| Authors | Audio-Visual features | | Method used for liveness detection | Database used | EER(%) |
|---|---|---|---|---|---|
| | Audio | Visual | | | |
| Chetty *et al.* [29] | MFCC | Geometric lip parameters | Gaussian mixture model | VidTIMIT | 0.6 |
| Eveno *et al.* [44] | LPC | Geometric lip parameters | Canonical correlation analysis Co-inertial analysis | XM2VTS | 12.5 |
| Chetty *et al.* [30] | MFCC | Geometric lip parameters | GMM-UBM | VidTIMIT, UCBN, AVOZES | 0.61 |
| Rua *et al.* [114] | MFCC | Discrete cosine transform | Co-Inertia analysis Coupled Hidden Markov Models | BANCA | 2.61 |
| Bredin *et al.* [22] | MFCC | Discrete cosine transform | Co-Inertia analysis | BANCA | 6.6 |
| Chetty [28] | MFCC | Multi channel gradient model | Reliability Weighted Summation | VidTIMIT | 8.6 |
| Zhu *et al.* [147] | MFCC | Discrete cosine transform | Co-Inertia Analysis with time delay | VidTIMIT | 14.6 |
| Boutella *et al.* [20] | MFCC | Space-time auto-correlation of gradients | Canonical correlation analysis, Partial least square analysis | BANCA, XM2VTS | 5.6 |

#### 2) FACE RECOGNITION UNDER VARYING ILLUMINATION CONDITIONS

Many researchers have used the XM2VTS [88], VidTIMIT [118] databases for implementing AV recognition systems. These databases are recorded in different sessions using different devices, but they do not contain biometric data with varying illuminations. The effect of varying illumination is among several bottlenecks in face recognition research. The proposed approaches using these databases might not perform well when the lighting changes are introduced. This problem is discussed in MOBIO [84] database, which contains samples with varying illuminations. So there is a need to include this dependency in AV biometric methods with a wide variety of illumination changes.

#### 3) USAGE OF ADVANCED SENSORS IN DATA CAPTURE

Advancements in sensor technology and computer vision have made it possible to capture images at multiple wavelengths. Most of the visual sensors capture visible wavelengths of light (e.g., RGB images). Multi-spectral sensors can capture wavelengths that spread at different wavelengths of the spectrum. Some advanced sensors can even capture signals at near-infrared radiation, short-wave radiations, and infrared radiation. The available literature and databases for traditional face recognition methods are based on the visible spectrum and face problems like pose variations and illumination changes. It has been proved that near-infrared image capturing improves face recognition performance. Therefore, it will be beneficial to use multi-spectral sensors in AV biometrics to overcome the problems in limited visible spectrum wavelengths.

#### 4) MULTI-LINGUAL SPEAKER RECOGNITION

The dependency of speaker recognition on the speaker's language has been observed in the recent works [82]. The mismatch of languages of speech samples in training, enrolling, and testing is a challenging problem in AV biometrics. Therefore, a multi-lingual AV biometric system is required for active research on this problem. A multi-lingual speaker recognition system aims to recognize a person based on speech features independent of language. It is observed that there are no previous works on AV biometrics performed the task of a multi-lingual speaker recognition system. So there is a broad scope for including the language dependency experiments in AV biometrics where the problem with language can be overcome with the complementing visual part.

### B. AV BIOMETRICS IN SMART DEVICES

Smartphone usage has grown from essential communication to multipurpose usage in the past decade. The key features of recent smartphones include mobile transactions, digital ID, and sensitive multimedia data transfer. The critical information involved in smartphone functionality requires more secure access than passwords or key phrases. Biometrics have come in to play in order to provide higher security in smartphone applications. Major smartphone vendors have deployed biometric sensors and recognition systems into smartphones (e.g., Touch ID, Face ID). Due to high security and easy to use, many third-party applications use the in-built biometric technology (e.g., Apple Pay). Banking applications like the iMobile app from ICICI bank uses fingerprint or face recognition for secure login.[13] However, the variance in devices and capturing situations in mobile environments restrict advanced biometric recognition. AV biometrics can solve some of the problems faced by unimodal recognition systems and provide better security in smartphones. There are multiple challenges around efficient utilization of AV biometrics in smartphones. Smartphones comes with wide variety of cameras and microphones to record the biometric characteristics. Therefore, the recognition algorithm should comprehend the huge variance in capturing channels and also adaptable to new devices.

Biometrics have been used in Internet of Things (IoT) devices to provide easy and secure control of the devices. The IoT devices contain sensitive information and use biometric technologies to protect the privacy [37]. However, the biometrics sensors embedded into IoT devices come with challenges like complex architecture in IoT infrastructure. The diverse set of devices and applications in IoT requires question the security provided by biometrics based authentication. Minute vulnerability in biometrics can pose a severe threat to the critical functionality of IoT. AV biometrics can provide a solution to the problems by providing better security than unimodal methods.

### C. PRIVACY PRESERVING TECHNIQUES IN AV BIOMETRICS

Biometric characteristics are unique to the person and cannot be changed. If the biometric template is compromised, there is a breach in the privacy of the individual. So the template security is the most crucial part of the biometric system. For tackling privacy protection, template protection techniques can be classified into three main categories: i) cancelable biometrics [97] consists of intentional, repeatable distortions of the biometric signal which are irreversibly transformed, ii) cryptobiometrics [26] where a key is generated from cryptographic algorithms, iii) biometrics in the encrypted domain [9] where homomorphic encryption techniques are applied to protect the biometric data. There are ample amount of literature available for face images in cancelable biometrics and in cryptobiometrics [76] also in homomorphic encryption [43], [96], [115]. And also there are limited amount of literature available for speech in the cancelable biometrics [17], [92], [99], [129], [142] and in homomorphic encryption [93], [131]. The privacy-preserving techniques in AV biometrics have not received much attention in the audio-visual domain equivalent to other biometrics. To the best of our knowledge, the three above mentioned techniques are not addressed for AV biometrics. Hence this an open problem that can be addressed in the AV domain.

#### 1) DE-IDENTIFICATION IN AV BIOMETRIC SYSTEM
De-identification is defined as concealing or removing the identity of the person or replacing it with a surrogate personal identity to prevent indirect or direct identification of the person. De-identification is an important tool to protect privacy, one of the most important social and political issues of today's information society. Face de-identification in still images has a lot of literature available starting from 2000 [111] and with the introduction of deep learning, GAN-based generative models have become the benchmark for learning faces. Similarly, face de-identification in videos has also gained interest [52]. The other cue speech has also gained much interest in de-identification. Speech de-identification is mainly based on voice transformation. Voice transformation refers to modifications of the non-linguistic characteristics (voice quality, voice individuality) of a given utterance without affecting its textual content. There are lot of literature available for voice de-identification for both text-dependent [98] and text-independent case [91]. The de-identification in AV biometrics has not received much heed when compared to individual cues. Hence it is an open problem that can be addressed.

### D. DEEP NEURAL NETWORK (DNN) BASED RECOGNITION

There are only two papers [4], [5] using DNN based methods for AV biometric recognition system where [4] used two different databases, and the later used only one database. The advantages of single capture bi-modal systems like AV biometrics can be exploited with deep learning approaches by developing an intelligent system using synchronous information. Using synchronous features makes it easy to avoid the problems caused by lighting, channel noise, and some presentation attacks. Therefore, there is a open research scope for utilizing advanced deep neural network based methods to develop efficient AV biometric recognition systems.

### E. PERFORMANCE EVALUATION FOR AV BIOMETRICS

The AV biometric research brings two types of biometric modalities into a single system. There are performance evaluation standards for testing biometric systems by ISO/IEC [66]. The most common metrics used by AV systems are FAR, FRR and EER. However, some research works on AV biometrics used different performance evaluation methods. Similarly, there are standard metrics by NIST namely False positive identification rate (FPIR) and False Negative identification rate (FNIR) for biometric vendor technology evaluations like face (FRVT) [55], iris (IREX) [56] and fingerprint (FpVTE) [141]. It is necessary to have a common evaluation protocol used in all the works to compare and comment on various AV biometric research works. In the case of vulnerability assessment, the impostor attack presentation match rate (IAPMR) has been a standard metric from ISO/IEC [68]. However, it is observed that EER has been used in most of the liveness detection methods. Although it makes it easy to compare different works, the EER values do not explicitly show the attacks' vulnerability. Therefore, there is a strong requirement of performance evaluation protocols to test the entire biometric system and include the effect of each cue on the whole system.

---

[13]https://www.icicibank.com/mobile-banking/imobile.page?#toptitle

## VIII. CONCLUSION AND FUTURE WORKS

Biometrics based person recognition has been used in multiple domains ranging from smartphone access, mobile transactions to border control checks. The vulnerabilities in unimodal biometric systems (audio or face only) make authentication systems prone to attacks questioning biometric recognition systems' resilience. The audio-visual biometric recognition systems have evolved to overcome these problems. The AV biometric methods take advantage of the complementary information present in correlated biometric cues, face, and voice. Over the year, multiple research works focused on AV biometrics and proposed efficient person recognition approaches. This survey paper has discussed how two complementary cues can provide critical information for AV biometric person recognition system. At first, we have presented the introduction on AV biometrics, discussed how different they are from other multimodal systems, and concepts of an ISO standard biometric recognition system and presentation attacks. Later, we have classified the different types of features used in AV biometric systems in both audio and visual domains, indicating their importance, advantages, and disadvantages. We have described the different approaches of information fusion of the two modalities used in AV biometric systems. We reviewed several AV biometric recognition systems that appeared in the literature and presented their experimental results.} We then shifted our focus onto the different databases used in the AV recognition system and presented a detailed discussion of the devices and capturing methods used in each of them. A comprehensive table is presented listing the best performing algorithm on each database. Further, we presented the PAD algorithms on AV based biometric systems. We have studied different feature extraction methods used for liveness detection and discussed the performance of detecting replay attacks. AV biometrics is a hot topic of research, with many accomplishments and exciting opportunities for further research and development. Keeping this in mind, we have discussed the challenges with several open problems and mentioned possible future research directions. Overall, this article can serve as a quick reference for AV biometric recognition systems and related PAD algorithms for beginners and experts.

### A. FUTURE WORKS

The detailed study on AV biometrics pointed out the challenges and open problems in this field. To overcome the challenges and solve open problems, the possible future works in this direction are briefly mentioned as follows.

- A novel database of AV biometric data can be implemented, including multiple dimensions like multiple languages, sessions, devices, and presentation attacks.
- State-of-the-art algorithms can be developed for defying the dependencies and vulnerabilities in AV biometrics.
- The advantages of AV biometrics like the correlation between face and voice can be exploited exclusively to overcome the generalization problem. This leads to new paths like visual speech or talking face biometrics.
- The growth of smartphone applications for sensitive usage can make use of AV biometrics. This direction needs a research focus on implementing AV based person recognition in a mobile environment.
- The multimodal biometrics requires special attention in protecting the stored sensitive biometrics data.

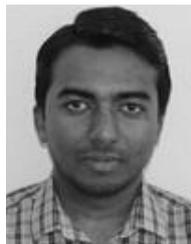

**HAREESH MANDALAPU** received the M.Tech. degree in computer science from the University of Hyderabad, in 2015, and the M.S. degree in Erasmus Masters CIMET from Université Jean Monnet, France, in 2017. He is currently pursuing the Ph.D. degree in information security and communication technology with the Norwegian University of Science and Technology, Gjøvik, Norway. His research interests include audio-visual biometrics, presentation attack detection, and multilingual speaker recognition.

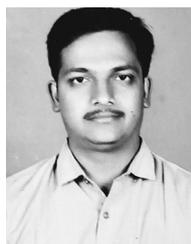

**ARAVINDA REDDY P N** received the M.Tech. degree in signal processing from Visvesvaraya Technological University Belgaum, in 2014. He is currently pursuing the Ph.D. degree with the Advanced Technology Development Centre, Indian Institute of Technology Kharagpur, Kharagpur, West Bengal, India. His research interests include automatic speech recognition, audio-visual biometrics, and presentation attack detection.

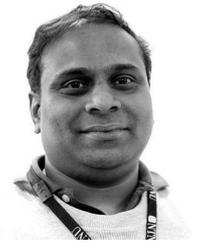

**RAGHAVENDRA RAMACHANDRA** (Senior Member, IEEE) received the Ph.D. degree in computer science and technology from the University of Mysore, Mysore, India, Institute Telecom, and Telecom Sudparis, Evry, France (carried out as a collaborative work), in 2010. He was a Researcher with the Istituto Italiano di Tecnologia, Genoa, Italy, where he worked with video surveillance and social signal processing. He is currently appointed as a Full Professor with the Institute of Information Security and communication technology (IIK), Norwegian University of Science and Technology (NTNU), Gjøvik, Norway. His main research interests include deep learning, statistical pattern recognition, data fusion schemes, and random optimization, with applications to biometrics, multimodal biometric fusion, human behaviour analysis, and crowd behaviour analysis. He has authored several papers and is a reviewer for several international conferences and journals. He also holds several patents in biometric presentation attack detection. He was/is also involved in various conference organizing and program committees and is serving as an Associate Editor for various journals. He was/is participating (as PI/Co-PI/contributor) in several EU projects, IARPA USA, and other national projects. He has served as an editor for ISO/IEC 24722 standards on multimodal biometrics and an active contributor for ISO/IEC SC 37 standards on biometrics. He has received several best paper awards.

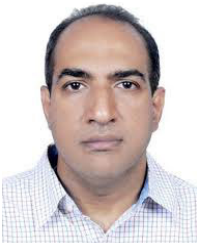

**KROTHAPALLI SREENIVASA RAO** (Member, IEEE) received the B.Tech. degree in electronics and communication from the RVR College of Engineering and the M.E. degree in communication systems from PSG Tech, Coimbatore, India, in 1990 and 2006, respectively, and the Ph.D. degree from the Department of Computer Science and Engineering, IIT Madras, Chennai, India, in 2004. He is currently working as a Professor with the Department of Computer Science and Engineering, IIT Kharagpur, Kharagpur, West Bengal, India. He has supervised seven PhDs and 14 MS (by research) in different issues related to speech processing.

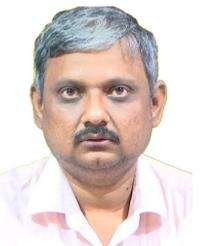

**PABITRA MITRA** (Member, IEEE) received the B.Tech. degree in electrical engineering from IIT Kharagpur, Kharagpur, India, in 1996, and the Ph.D. degree from the Department of Computer Science and Engineering, Indian Statistical Institute, Kolkata, India, in 2005. He is currently working as a Professor with the Department of Computer Science and Engineering, IIT Kharagpur. He has supervised eight PhDs and 12 MS (by research) in different issues related to AI and machine learning.

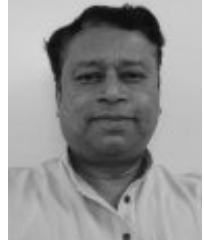

**S. R. MAHADEVA PRASANNA** (Member, IEEE) received the B.Tech. degree in electronics and communication from SSIT Tumakuru, Tumakuru, Karnataka, India, in 1994, the M.Tech. degree in industrial electronics from NIT Surathkal, Surathkal, Karnataka, in 1997, and the Ph.D. degree from the Department of Computer Science and Engineering, IIT Madras, Chennai, India, in 2004. He is currently working as a Professor with the Department of Electrical Engineering, IIT Dharwad, Dharwad, India. He has supervised 13 PhDs in different issues related to speech processing.

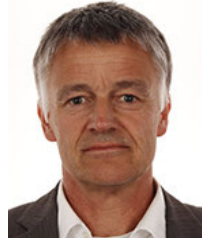

**CHRISTOPH BUSCH** (Senior Member, IEEE) is a member of the Department of Information Security and Communication Technology (IIK), Norwegian University of Science and Technology (NTNU), Norway. He holds a joint appointment with the Faculty of Computer Science, Hochschule Darmstadt (HDA), Germany. Furthermore, he has been a Lecturer of Biometric Systems with the Technical University of Denmark (DTU), since 2007. He has coauthored more than 400 technical papers and has been a speaker at international conferences. He is a Convener of WG3 in ISO/IEC JTC1 SC37 on biometrics and an active member of CEN TC 224 WG18. He has served for various program committees, such as NIST IBPC, ICB, ICHB, BSI-Congress, GI-Congress, DACH, WEDELMUSIC, and EUROGRAPHICS, and served for several conferences, journals, and magazines as a Reviewer such as ACM-SIGGRAPH, ACM-TISSEC, the IEEE COMPUTER GRAPHICS AND APPLICATIONS, the IEEE TRANSACTIONS ON SIGNAL PROCESSING, the IEEE TRANSACTIONS ON INFORMATION FORENSICS AND SECURITY, the IEEE TRANSACTIONS ON PATTERN ANALYSIS AND MACHINE INTELLIGENCE, and the *Computers and Security Journal* (Elsevier). Furthermore, on behalf of Fraunhofer, he chairs the biometrics working group of the TeleTrusT association as well as the German standardization body on biometrics (DIN-NIA37). He is also an Appointed Member of the Editorial Board of the *IET Biometrics journal* and the IEEE TRANSACTIONS ON INFORMATION FORENSICS AND SECURITY journal.

○ ○ ○